\documentclass[pre,showpacs,twocolumn,showkeys,amssymb,amsmath,aps]{revtex4}

\usepackage{graphicx}
\usepackage{bm}
\begin{document}
\title{Entropy reduction from a detailed fluctuation theorem for a
nonequilibrium stochastic system driven under feedback control}
\author{M. Ponmurugan\footnote{email: mpn@imsc.res.in}} 
\affiliation{The Institute of Mathematical Sciences, 
C.I.T. Campus, Taramani, Chennai 600113, India}
\date{\today}
\begin{abstract} 
We show explicitly the entropy reduction from a detailed fluctuation 
theorem  for the general stochastic 
system driven by nonequilibrium process under feedback control.
The effect of interaction of the feedback controller with 
the system is to reduce the entropy of the system.  
We define the entropy reduction for the single trajectory 
and show that the overall entropy production for the stochastic 
system with feedback control can be either positive or negative. 
The negative entropy production  has been 
studied earlier for a simple system with velocity dependent feedback control 
[K. H. Kim and H. Qian, Phys. Rev. Lett. {\bf 93}, 120602 (2004)]. 
Our general approach  provides the overall 
positive or negative entropy production irrespective of 
velocity dependent and position dependent feedback control.
\end{abstract}
\maketitle

\section{Introduction}
The evolution of the physical systems can be modified or controlled 
by repeated operation of an external agent called controller \cite{frev,feedref}.
The action of the controller is to regulate the system 
dynamics and increase its performance. The controller 
can operate on the system blindly or it can use 
information about the state of the system. The former is
known as the open loop controller and the latter one is called 
as feedback or closed loop controller. 
The feedback controller measures the 
partial performance of the system, and its action 
on the system depends on the outcome of measurements \cite{cao}.
Recent advances in nanotechnology allows the active control of 
the position and velocity of the single molecule by a feedback 
system \cite{na1,qian}. The proper utilization of the information about the 
state of the system in feedback control effectively 
improves the system performance \cite{feedref,frev,cao}.

In last two decades, the concept of fluctuation theorems has
become an active research area in statistical physics 
for advanced theoretical understanding and 
experimental verifications \cite{other,col}. 
This theorem generally establishes the 
connections between the nonequilibrium
stochastic fluctuations of system and 
its dissipative properties.
Further it establishes rigorous identities for the 
nonequilibrium averages of thermodynamic observables 
such as work, heat, entropy, or current \cite{jar,crooks,seifert}.

Consider a system initially
in equilibrium at temperature (inverse) $\beta=1/k_BT$
($k_B$ is the Boltzmann constant) which is externally driven from 
its initial state to final state by nonequilibrium process.
Let $P[\Gamma(t)]$ be the
probability of the phase space trajectory, $\Gamma(t)$
for the system driven between 
the two states in forward direction.
This satisfies the detailed fluctuation theorem \cite{crooks,seifert,horo,sagaf},
\begin{eqnarray}\label{fluc}
  \frac{P[\Gamma(t)]}{P^{\dagger}[\Gamma^{\dagger}(t)]} &=& e^{\sigma[\Gamma(t)]},
\end{eqnarray}
where $\sigma[\Gamma(t)]$ is the positive entropy 
production of the driven system and $P^{\dagger}[\Gamma^{\dagger}(t)]$ is the
probability of the phase space trajectory $\Gamma^{\dagger}(t)$,
for the system driven in the  reversed direction.
This is the direct relation 
between the entropy production and the ratio of probabilities for
the forward and the reversed trajectories \cite{seifert,horo}.

The presence of feedback control in a physical system  in general modifies both  
the fluctuation theorem and the nonequilibrium averages  
such as Jarzynski equality \cite{qian,sagaf}. 
We have shown recently that the feedback control in a physical 
system should preserve the detailed fluctuation theorem if 
the system has the same feedback information 
measure in both directions  \cite{ponf1}. 
Our results {are} based on the assumption that the system with 
feedback control should locally satisfy the 
detailed fluctuation theorem.
Such an assumption leads to a positive entropy 
production from a detailed fluctuation theorem for 
the system with feedback control. 
However, the effect of the interaction of the controller with the 
system is to reduce the entropy of the system \cite{cao,qian} which 
was not captured in our earlier results \cite{ponf1}.

The entropy production can be positive 
or negative depending on the feedback control.
The negative entropy production can happen generally in a 
system with velocity dependent feedback control (VFC) such as 
molecular refrigerator \cite{qian}. There is a general belief that 
the entropy reduction in velocity dependent feedback control (VFC) is
significantly different from the stochastic 
systems with position dependent feedback control (PFC) \cite{qian}.
Since the entropy reduction in the system due to the information used by 
the feedback controller is quite general for both VFC and PFC,
it will be worth to discuss the entropy reduction from a  fluctuation 
theorem due to feedback control in a general framework.

In following a few works on the single trajectory entropy  
for the stochastic system \cite{crooks,seifert,qian1}, in this paper,
we define for the first time the entropy reduction of the single  trajectory,   
and show explicitly the entropy reduction from a detailed fluctuation theorem 
for the general stochastic system  driven under feedback control. 
The explicit appearance of the entropy reduction term in fluctuation theorem 
is an important result of this work. In addition, 
our approach allows one to compute the overall positive 
or negative entropy production irrespective of PFC or VFC.

\section{Single trajectory entropy reduction due to feedback control}

The feedback control enhances our controllability of 
small thermodynamic systems \cite{frev,feedref,cao}. 
Whenever the controller measures the 
partial state of the system, the result of 
the measurement determines the action of the controller. 
The additional information on the system 
provided by the measurement further 
determines the state of the system.
Suppose, the controller performs a measurement on a  stochastic 
thermodynamic system at time $t_m$. Let $\Gamma_m$ be the 
phase-space point of the system at that time,
$P[\Gamma_m]$ its probability, and $y$ the measurement.
Depending on the controller measurements,  
the outcome $y$ can occur 
with a probability $P[y]$ \cite{sagaf}.
The information obtained by the controller can be characterized
by the mutual (feedback) information measure \cite{sagaf,ponf1},
\begin{eqnarray}\label{eid}
I[y,\Gamma_m]=ln \left[ \frac{P[y|\Gamma_m]}{P[y]} \right ],
\end{eqnarray}
where $P[y|\Gamma_m]$ is the conditional probability of 
obtaining outcome $y$ on condition that the state of the system 
is $\Gamma_m$.

In experiments and simulations, the path connecting the 
two states of the system in the time period $\tau$ can be obtained 
by pulling the system from one state
to another along a switching path. This can be parameterized  
using the switching protocol path variable $\lambda$ \cite{jar,sagaf}.
The switching rate determines whether the switching process 
is an equilibrium (infinitely slow) or nonequilibrium (fast) process.
If the experiments are performed under feedback control,
the switching control parameter $\lambda$ depends on the outcome 
$y$ after $t_m$ \cite{sagaf}.
That is, whenever the controller makes measurements, there 
is a corresponding change in the switching parameter
for the next time step, which is denoted as $\lambda_{(t;y)}$.
After each measurement, 
the value of outcome $y$ should be fixed and the corresponding 
switching parameter $\lambda_{(t;y)}$ does not change 
until the controller make the another measurement.

Let $P_{\lambda_{(t;y)}}[\Gamma(t)]$ be the 
probability of the phase space trajectory, $\Gamma(t)$
in forward direction for the   
switching protocol $\lambda_{(t;y)}$. 
Since the entropy of the system is reduced by feedback 
control \cite{cao}, we define the single trajectory 
entropy reduction for the protocol $\lambda_{(t;y)}$ as 
\begin{eqnarray}\label{reddefwer}
\sigma_{r}[\Gamma(t)]  &=& ln
\left[ \frac{P_{\lambda^{\star}}[\Gamma(t)]}{P_{\lambda_{(t;y)}}[\Gamma(t)]} \right].
\end{eqnarray}
Here $P_{\lambda^{\star}}[\Gamma(t)]$ is the probability of the 
phase space trajectory for the system 
without feedback control driven by an arbitrary chosen   
switching protocol ${\lambda^{\star}}$ (see, Eq.(\ref{fluc})).
It should be noted that prior to the controller measurement the feedback experiment
initially started  with the same protocol ${\lambda^{\star}}$ . 
Our definition of entropy reduction has a definite meaning in the sense that
if the system  has no effect of feedback control,
$P_{\lambda_{(t;y)}}[\Gamma(t)]=P_{\lambda^{\star}}[\Gamma(t)]$,
which results in zero entropy reduction.
However, one can see latter from  Eq.(\ref{rent})
that there should be a single trajectory entropy production. 
If the driven system strongly influenced by the feedback controller,
$P_{\lambda_{(t;y)}}[\Gamma(t)]$ should be significantly  different
from $P_{\lambda^{\star}}[\Gamma(t)]$, which eventually results 
in reduction in entropy \cite{cao}.

Starting from the same switching protocol $\lambda^{\star}$ 
one can perform the feedback control experiment in
reverse direction by driving the system from 
the final equilibrium state of the forward switching process 
to its initial equilibrium state. 
Let $P^{\dagger}_{\lambda^{\star}}[\Gamma^{\dagger}(t)]$ be the 
probability of the phase space trajectory, $\Gamma^{\dagger}(t)$ 
of the system without feedback control driven in the reverse direction 
for the switching protocol $\lambda^{\star}$ (see, Eq.(\ref{fluc})).
For every trajectory in the forward direction there should be a 
corresponding trajectory in the reverse direction
\cite{crooks,seifert,horo,sagaf}. 
If we perform the feedback controller 
measurements for a given system in the reverse direction,
the single trajectory entropy reduction in the reverse 
direction is  
\begin{eqnarray}\label{revent}
\sigma^{\dagger}_{r}[\Gamma^{\dagger}(t)] &=&
-\sigma_{r}[\Gamma(t)], 
\end{eqnarray}
From the definition of Eq.(\ref{reddefwer}),  the single trajectory
entropy reduction in reverse direction is given by 
\begin{eqnarray}\label{reddefwer1}
\sigma^{\dagger}_{r}[\Gamma^{\dagger}(t)] &=& 
ln \left[ \frac{P^{\dagger}_{\lambda^{\star}}[\Gamma^{\dagger}(t)]}{P^{\dagger}_{\lambda_{(t;y^{\dagger})}^{\dagger}}[\Gamma^{\dagger}(t)]} \right],
\end{eqnarray}
where $P^{\dagger}_{\lambda_{(t;y^{\dagger})}^{\dagger}}[\Gamma^{\dagger}(t)]$
is the probability of the phase space trajectory
in reverse direction for the switching protocol $\lambda_{(t;y^{\dagger})}^{\dagger}$.
The measurement outcome $y^{\dagger}$  can  
occur with probability $P^{\dagger}[y^{\dagger}]$. 
The mutual information 
measure due to feedback control in reverse direction is given by 
\begin{eqnarray}\label{eidr}
I^{\dagger}[y^{\dagger},\Gamma^{\dagger}_m]=ln \left[ \frac{P^{\dagger}[y^{\dagger}|\Gamma^{\dagger}_m]}{P^{\dagger}[y^{\dagger}]} \right ],
\end{eqnarray}
where $P^{\dagger}[y^{\dagger}|\Gamma^{\dagger}_m]$ is the conditional probability of 
obtaining outcome $y^{\dagger}$ on the condition that the state of the system 
is $\Gamma^{\dagger}_m$.

Using Eq.(\ref{fluc}), Eq.(\ref{reddefwer1}) can be rewritten as
\begin{eqnarray}
\sigma^{\dagger}_{r}[\Gamma^{\dagger}(t)]  &=&  
 ln \left[ \frac{P^{\dagger}_{\lambda^{\star}}[\Gamma^{\dagger}(t)]}{P_{\lambda^{\star}}[\Gamma(t)]} \right]  \nonumber \\
&+& ln \left[ \frac{P_{\lambda^{\star}}[\Gamma(t)]}{P^{\dagger}_{\lambda_{(t;y^{\dagger})}^{\dagger}}[\Gamma^{\dagger}(t)]} \right]
\nonumber \\
&=& - \sigma[\Gamma(t)] +
ln \left[ \frac{P_{\lambda^{\star}}[\Gamma(t)]}{P^{\dagger}_{\lambda_{(t;y^{\dagger})}^{\dagger}}[\Gamma^{\dagger}(t)]} \right]. \nonumber 
\end{eqnarray}
Therefore,
\begin{eqnarray}
ln \left[ \frac{P_{\lambda^{\star}}[\Gamma(t)]}{P^{\dagger}_{\lambda_{(t;y^{\dagger})}^{\dagger}}[\Gamma^{\dagger}(t)]} \right] &=&
\sigma^{\dagger}_{r}[\Gamma^{\dagger}(t)] +
\sigma[\Gamma(t)]. \nonumber \\
\end{eqnarray}
The left hand side of the above equation can be rewritten as
\begin{eqnarray}
ln \left[ \frac{P_{\lambda^{\star}}[\Gamma(t)]} {P_{\lambda_{(t;y)}}[\Gamma(t)]} \right] + 
ln \left[ \frac{P_{\lambda_{(t;y)}}[\Gamma(t)]} {P^{\dagger}_{\lambda_{(t;y^{\dagger})}^{\dagger}}[\Gamma^{\dagger}(t)]} \right]
 &=& \nonumber \\
\sigma^{\dagger}_{r}[\Gamma^{\dagger}(t)] + \sigma[\Gamma(t)].
\end{eqnarray}
Using Eq.(\ref{reddefwer}) and Eq.(\ref{revent}), the above equation becomes,
\begin{eqnarray}\label{locfluc}
\frac{P_{\lambda_{(t;y)}}[\Gamma(t)]} {P^{\dagger}_{\lambda_{(t;y^{\dagger})}^{\dagger}}[\Gamma^{\dagger}(t)]} =e^{S_{r}[\Gamma(t)]},
\end{eqnarray}
where
\begin{eqnarray}\label{rent} 
S_{r}[\Gamma(t)] &=& -2 \sigma_{r}[\Gamma(t)] + \sigma[\Gamma(t)]
\end{eqnarray}
whose reversed part 
\begin{eqnarray}\label{revent1}
S^{\dagger}_{r}[\Gamma^{\dagger}(t)]&=& - S_{r}[\Gamma(t)].
\end{eqnarray}

It is clear from Eq.(\ref{rent}) that 
if the feedback control has no effect on the system,
$\sigma_{r}[\Gamma(t)]=0$, however, there should be 
a single trajectory entropy production 
$S_{r}[\Gamma(t)] = \sigma[\Gamma(t)]$.

Under identical experimental conditions in the forward and 
reverse directions, we assume that the error in the controller 
measurements outcomes $y$ and $y^{\dagger}$ should be the same. 
In such a case,
\begin{eqnarray}\label{iequal}
P[y]&=& P^{\dagger}[y^{\dagger}]. 
\end{eqnarray}

Combining Eq.(\ref{eid}), Eq.(\ref{eidr}) and Eq.(\ref{iequal}){,} 
we can obtain 
\begin{eqnarray}\label{iratio}
\frac{P[y|\Gamma_m]}{P^{\dagger}[y^{\dagger}|\Gamma^{\dagger}_m]} &=&
e^{I[y,\Gamma_m]-I^{\dagger}[y^{\dagger},\Gamma^{\dagger}_m]}.
\end{eqnarray}
For the controller measurement condition 
$I[y,\Gamma_m]=I^{\dagger}[y^{\dagger},\Gamma^{\dagger}_m]$ \cite{ponf1},
the above equation becomes,
\begin{eqnarray}\label{irat}
P[y|\Gamma_m]=P^{\dagger}[y^{\dagger}|\Gamma^{\dagger}_m]. 
\end{eqnarray}
If we obtained the same measurement 
outcome $y'$ \cite{sagaf,ponf1} 
of phase point $\Gamma'$ in the forward direction and 
$\Gamma'^{\dagger}$ in the reverse direction then 
Eq.(\ref{locfluc}) and Eq.(\ref{irat}) become,
\begin{eqnarray}\label{locflucp}
\frac{P_{\lambda_{(t;y')}}[\Gamma(t)]} {P^{\dagger}_{\lambda_{(t;y')}^{\dagger}}[\Gamma^{\dagger}(t)]} =e^{S_{r}[\Gamma(t)]}
\end{eqnarray}
\begin{eqnarray}\label{iratio2}
P[y'|\Gamma']=P^{\dagger}[y'|\Gamma'^{\dagger}]. 
\end{eqnarray}
In what follows we use the above equations and obtain the  
overall entropy production from a detailed fluctuation 
theorem due to feedback control
in both the directions.  

\section{Detailed fluctuation theorem under feedback control}

Let $P[\widetilde{S}_r]$ be 
the probability of obtaining $S_r[\widetilde{\Gamma}]$ 
from the repeated feedback control experiment in 
the forward direction. 
This probability can be obtained from the 
joint distribution of $\Gamma(t)$ and $y$ is given 
by \cite{sagaf,ponf1} 
\begin{eqnarray}\label{fl1}
P[\widetilde{S}_r] &=& \int  P[y'|\Gamma'] P_{\lambda_{(t;y')}}[\Gamma(t)]\nonumber \\ 
  & &\delta(S_r[\Gamma(t)]-S_r[\widetilde{\Gamma}]) \ dy' \ {\it D}[\Gamma(t)],
\end{eqnarray}
where $\delta(x)$ is the Dirac delta function which has a property
$\delta(-x)=\delta(x)$. Using Eq.(\ref{locflucp}), the 
above equation becomes, 
\begin{eqnarray}
P[\widetilde{S}_r] &=& \int e^{S_r[\Gamma(t)]} P[y'|\Gamma'] P^{\dagger}_{\lambda_{(t;y')}^{\dagger}}[\Gamma^{\dagger}(t)] \nonumber \\
& & \delta(S_r[\Gamma(t)]-S_r[\widetilde{\Gamma}]) dy' \ {\it D}[\Gamma(t)], \nonumber
\end{eqnarray}

\begin{eqnarray}\label{fl2}
P[\widetilde{S}_r] &=& e^{S_r[\widetilde{\Gamma}]} \int  P[y'|\Gamma']  P^{\dagger}_{\lambda_{(t;y')}^{\dagger}}[\Gamma^{\dagger}(t)] \nonumber \\
& & \delta(S_r[\Gamma(t)]-S_r[\widetilde{\Gamma}]) dy' \ {\it D}[\Gamma(t)], \nonumber
\end{eqnarray}
Using Eq.(\ref{revent1}) and  Eq.(\ref{iratio2}){,} the above integral can be rewritten as,
\begin{eqnarray}
P[\widetilde{S}_r] &=& e^{S_r[\widetilde{\Gamma}]} 
\int P^{\dagger}[y'|\Gamma'^{\dagger}]  P^{\dagger}_{\lambda_{(t;y')}^{\dagger}}[\Gamma^{\dagger}(t)] \nonumber \\
& & \delta(S^{\dagger}_r[\Gamma^{\dagger}(t)]-S^{\dagger}_r[\widetilde{\Gamma}^{\dagger}]) \nonumber \\
& & dy' \ {\it D}[\Gamma^{\dagger}(t)].
\end{eqnarray}
Since $\delta(-x)=\delta(x)$ and ${\it D}[\Gamma^{\dagger}(t)]={\it D}[\Gamma(t)]$ \cite{sagaf},
we can obtain the overall entropy production from a detailed fluctuation theorem under 
feedback control in both directions as
\begin{eqnarray}\label{main}
\frac{P[\widetilde{S}_r]} {P^{\dagger}[-\widetilde{S}_r]}  &=& e^{S_r[\widetilde{\Gamma}]}, 
\end{eqnarray}
where $P^{\dagger}[-\widetilde{S}_r]$ is  the 
probability of obtaining $S^{\dagger}_r[\widetilde{\Gamma}^{\dagger}]$ 
from the repeated feedback control experiment in the reverse direction. 
This result explicitly provides the entropy reduction in the controlled 
system due to the external agent that 
operates on it. We can rewrite Eq.(\ref{main}) simply as
\begin{eqnarray}\label{main1}
\frac{P[-2 \sigma_{r} + \sigma]}{P^{\dagger}[-(-2 \sigma_{r} + \sigma)]}
&=& e^{-2\sigma_{r} + \sigma}.
\end{eqnarray}

From the above relation, we can discuss the different 
entropy reduction conditions as follows.
If $\sigma_r < \sigma/2$, $S_r$ is always positive, which 
provides the overall positive entropy production.
It is clear that the positive entropy production 
as  obtained in our earlier work \cite{ponf1} for the system
with feedback control in both directions 
is a special case of the present derivation. 
If $\sigma_r = \sigma/2$, the  overall entropy production is zero. This 
may be a condition for equilibrium \cite{qian}.
If $\sigma_r > \sigma/2$, $S_r$ is always negative,
which provides the overall negative entropy production.
This situation can happen quite generally in velocity dependent 
feedback control experiment (molecular refrigerator) \cite{qian}.
This shows that our  general result is valid for both the velocity dependent and
position dependent feedback control. The overall entropy production can be 
positive or negative depending upon the interaction of the system 
with the feedback controller.

Finally, using Eq.(\ref{main1}), it is straightforward to obtain 
the integral fluctuation theorem \cite{seifert}, 
\begin{eqnarray}
\langle e^{-(-2\sigma_{r} + \sigma)} \rangle = 1.
\end{eqnarray}
If the feedback control has no effect on the system,
$\sigma_r=0$, then we get positive 
entropy production detailed fluctuation theorem \cite{com}
\begin{eqnarray}
\frac{P[\sigma]}{P^{\dagger}[-\sigma]}= e^{\sigma}.
\end{eqnarray}

\section{Conclusion}

We have defined the single trajectory entropy reduction 
for the nonequilibrium stochastic system 
driven under feedback control.
Our result (Eq.\ref{main1}) explicitly shows the 
entropy reduction from a detailed fluctuation theorem for a
stochastic system driven under feedback control.
Our general result is valid for both the velocity dependent and 
the position dependent feedback control. This  
allows one to compute the overall entropy production
which can be either positive or negative. 

\paragraph*{\bf Acknowledgement:}
I would like to thank Renugambal Purushothaman for proof-reading the manuscript.


\begin{thebibliography}{100}
\bibitem{com}
Based on the information about the forward switching 
protocols, without feedback control,
one can perform the same variable switching experiment 
in {the} reverse direction  and can obtain the
extended detailed fluctuation theorem \cite{ponf1,vaikunt}.
Given below is a reply to a comment of Ref.\cite{vaikunt}  
on the system with feedback control
in forward direction alone\cite{ponf1}.
If there is no feedback in reverse direction, 
we stress that the mutual information measure, 
$I^{\dagger}[y^{\dagger},\Gamma^{\dagger}_m]=0$ \cite{sgcom}
in reverse direction.
However, the system driven in reverse direction 
should contain the same experimental error information
as in the forward direction. Since 
the source of experimental error is same in both the directions,
one can obtain from  Eq.(\ref{iratio}) that   
\begin{eqnarray}
I[y,\Gamma_m]= ln \left[\frac{P[y|\Gamma_m]}{P^{\dagger}[y^{\dagger}|\Gamma^{\dagger}_m]}\right].\nonumber
\end{eqnarray}
This relation is similar to the  
change of uncertainty as defined in Ref.\cite{vaikunt}
(see, cf Eq.(15) of Ref.\cite{vaikunt} for single loop feedback).
\end{thebibliography}
\end{document}